\documentclass[aps,prd,reprint,nofootinbib,citeautoscript,floatfix]{revtex4-1}

\newcommand{\LorP}{letter}

\usepackage{amsmath}
\usepackage[export]{adjustbox}
\usepackage{color}
\usepackage{graphicx}
\usepackage{dcolumn}
\usepackage{bm}
\usepackage{hyperref}
\usepackage{natbib}
\usepackage{float}
\usepackage[caption=false]{subfig}
\usepackage{subfloat}
\usepackage{hyperref}
\hypersetup{
colorlinks=true,
linktoc=page,    
linkcolor=blue,
citecolor=blue,
urlcolor=blue,
colorlinks=true,
}

\newcommand{\be}{\begin{equation}}
\newcommand{\ee}{\end{equation}}
\newcommand{\bea}{\begin{eqnarray}}
\newcommand{\eea}{\end{eqnarray}}

\newcommand{\mH}{\mathcal{H}}

\newcommand{\mO}{\mathcal{O}}

\newcommand{\mR}{\mathcal{R}}

\newcommand{\tr}{\text{tr}}

\def\/{\frac}

\newcommand{\nn}{\nonumber}

\newcommand{\bra}[1]{\left<{#1}\right|}

\newcommand{\ket}[1]{\left|{#1}\right>}

\begin{document}

\title{Evaporating black holes and late-stage loss of soft hair
}

\author{Peng Cheng$^{1, 2}$}\email{p.cheng.nl@outlook.com}
\affiliation{1) Center for Joint Quantum Studies and Department of Physics, Tianjin University, 300350 Tianjin, P. R. China\\
2) Institute for Theoretical Physics, University of Amsterdam, 1090 GL Amsterdam, The Netherlands
}

\begin{abstract}
We present a paradox for evaporating black holes, which is common in most schemes trying to avoid the firewall by decoupling early and late radiation. At the late stage of the black hole evaporation, the decoupling between early and late radiation can not be realized because the black hole has a very small coarse-grained entropy, then we are faced with the firewall again. We call the problem hair-loss paradox as a pun on losing black hole soft hair during the black hole evaporation and 
the situation that the information paradox has put so much pressure on researchers.
\end{abstract}

\maketitle
\flushbottom

\section{Introduction}
\label{intro}

Black holes are fascinating objects in the universe, which have brought us many theoretical puzzles at the same time. The black hole information paradox (BHIP), as an example, was firstly put forward by Stephen Hawking \cite{Hawking1976} and has been confusing researchers for almost half a century now.

As a milestone, the BHIP was sharpened by the Almheiri-Marolf-Polchinski-Sully (AMPS) firewall argument \cite{Almheiri2012} in 2012. The basic idea can be summarized as follows. 
For an old black hole, we can denote the early radiation as $R$ and the partner of a late time Hawking radiation $H$ as $P$. The Penrose diagran is shown in Fig. \ref{penrose}. The unitarity requires that $H$ purifies the early radiation $R$, such that we have the Page curve \cite{Page1993,Page2013} for the entropy of the overall radiation. Let us use $R_H$ to denote the part of $R$ that is maximally entangled with the Hawking radiation $H$. Furthermore, the smoothness of the horizon requires that $P$ and $H$ are maximally entangled. However, the unitarity and the smoothness of the horizon are not consistent with each other because of the so-called \textit{monogamy of entanglement} \cite{Terhal2004}, which claims that $H$ can only be entangled with either $R$ or $P$. Then, the unitarity implies the horizon is not smooth anymore.

The Harlow-Hayden-Aaronson (HHA) decoding task \cite{Harlow2013,Aaronson2016} was proposed as a possible solution to avoid the firewall.
The basic logic is very similar to the black hole complementarity \cite{Susskind_1993,Kiem1995,Lowe1995}. 
We said that $H$ is maximally entangled with $R_H$, which is only a small part of $R$. As far as no one can distill $R_H$ out of $R$, we can think $H$ is decoupled with the early radiation $R$. If $H$ and $R$ are decoupled, the firewall can be avoided.
The quantum computation argument claims that the task of distilling $R_H$ out of $R$ is as hard
as inverting a one-way function, and the distilling time can be much longer than the evaporating time of the black hole. 

Although the decoding task provided a nice quantum computational argument, questions like how to physically realize the decoding task and the derivation of the Page curve was not addressed convincingly until very recently. 
Yoshida considered the back-reaction of an infalling observer $O$ in solving the black hole complexity puzzle and soon realized it could also be used to achieve the decoupling between $H$ and $R$ \cite{Yoshida2019}, as shown in Fig. \ref{recover}.
The basic idea can be summarized as Yoshida's decoupling theorem \cite{Yoshida2017,Yoshida2019}: for scrambling black hole evolution, if the dimension of the observer's Hilbert space $\mH_O$ is much larger than the late time Hawking radiation $\mH_H$, i.e., $d_O\gg d_H$, early and late radiation are decoupled, and one can reconstruct the Hawking partner $P$ without using the early radiation $R$.
The inclusion of observer Hilbert space was legitimized by gravitational dressing, and the observer degrees of freedom can be identified with the soft hair degrees of freedom $\mH_S$ of a black hole system \cite{Hosur2016,Yoshida2019a,Pasterski2020,Li2021a}. Here we are going to assume the entropy of the soft hair contributes a large portion of the black hole entropy.
It has also been shown in \cite{Pasterski2020} that the soft hair introduces an observer-dependent firewall, and the construction of $P$ can be performed via the Petz map \cite{Petz1988,DenesPetz2004}.
  The Page curve was also addressed in \cite{Cheng2020}.

However, the current letter is trying to emphasize that the above argument relies heavily on the dimension of the soft hair Hilbert space $d_S$ is much larger than the Hilbert space of $H$, i.e. 
\be d_S=d_O\gg d_H\,.\ee
Furthermore, the error of reconstructing the interior observer is determined by $d_H/d_S$.
However, the above condition is not fulfilled at the end of the evaporation. The reason is as follows. The coarse-grained entropy of the black hole decreases as the black hole evaporates, and the entropy of the soft hair can not be large than the coarse-grained entropy of the black hole. So when the black hole entropy is relatively small, $d_S$ being much larger than $d_H$ can not be fullfilled.
Then, the decoupling theorem fails at the end of the evaporation, and the firewall shows up again.
If the firewall shows up once, the whole horizon would be ``lighted", and there would be a firewall all over the horizon because of the general covariance of the general relativity. 
We call the problem \textit{hair-loss paradox}, as a pun on the annoyance of losing black hole ``soft hair".
The hair-loss paradox is actually a problem for most of the arguments trying to decouple $R$ and $H$. In the HHA decoding task, if $R_H$ is only left with a few qubits, then the distillation of $R_H$ out of $R$ is not that hard anymore, and we would have to face with the firewall again.

Heuristically, the decoupling can always be achieved if there is a large extra system.  If the entanglement between the early and late radiation is like looking for a needle in the ocean, we may just ignore it. However, at the very end of the evaporation, the entropies of the black hole and radiation are both very small, so there would be not possible to have a large extra system to realize the decoupling between early and late radiation. It is like if the ocean evaporates to a few drops of water, the needle cannot be ignored anymore. So, even if the firewall paradox can be fixed near the Page time, we always end up with a firewall at the end of the evaporation.

\section{Soft hair in the BHIP}
\label{review}

In this section, we review the basic ideas of the decoupling theorem and the role of soft hair in the BHIP. The review mainly follows \cite{Pasterski2020} (also see \cite{Cheng2020}). 
The decoupling between $R$ and $H$ can be realized by including a large exterior observer $O$, and the measurement of soft hair by the observer can be used to reconstruct the Hawking partner $P$ via the Petz map.  
\begin{figure}[H]
  \centering
\includegraphics[width=5cm]{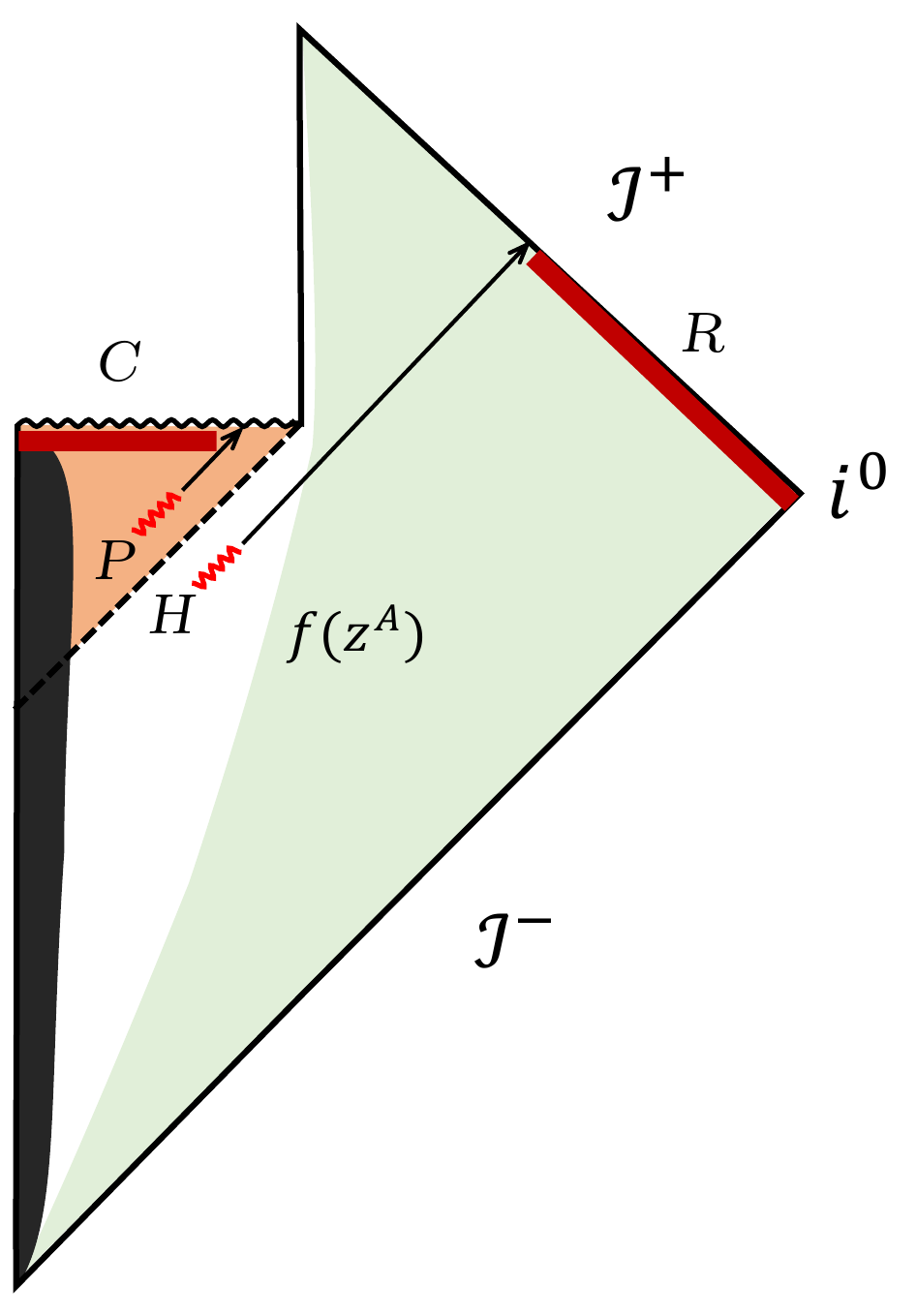}
  \caption{ Penrose diagram of an evaporating black hole. $R$ is the early radiation collected at infinity; $C$ represents the particles collected in the interior of the black hole. After a short period of time, the black hole emits an extra Hawking radiation $H$, and its partner is represented by $P$. The correlation between black hole soft hair $S$ and exterior observer $O$ is encoded in a function $f(z^A)$. 
  }\label{penrose}
  \end{figure}
First of all, let us describe the correlation between the exterior operator $\mO$ and the black hole via gravitational dressing.
Any diffeomorphism invariant operator must commute with the supertransition charge $Q_f$, i.e.
\be
[Q_f,\mO_{phys}]=0\,.
\ee
The above requirement enforces that the physical operator is dressed with gravitational Wilson line \cite{Himwich2020}
\be
\mathcal{W}(\omega,z^A)=e^{-i\omega f(z^A)}\,,
\ee
where $z^A$ are the coordinates on the celestial sphere \cite{Strominger2017}, and $\omega$ is the energy eigenvalue. So the gravitational dressing amount to add an extra phase $f$ along the light-cone coordinate $v'=v-f$. In other words, in order to be physical, all operators should know the supertransition charge of the spacetime via gravitational dressing. In an asymptotic flat black hole background, the transition function $f$ relating the near-horizon region and the asymptotic region can be interpreted as the soft hair of the black hole, which can have a large contribution to the black hole entropy. In such a sense, exterior observer has correlation with the black hole via soft hair. The soft theorems in curved spacetime is further investigated in \cite{Donnay2016,Hawking2016,Cheng2022}.

\begin{figure}[H]
  \centering
\includegraphics[width=8cm]{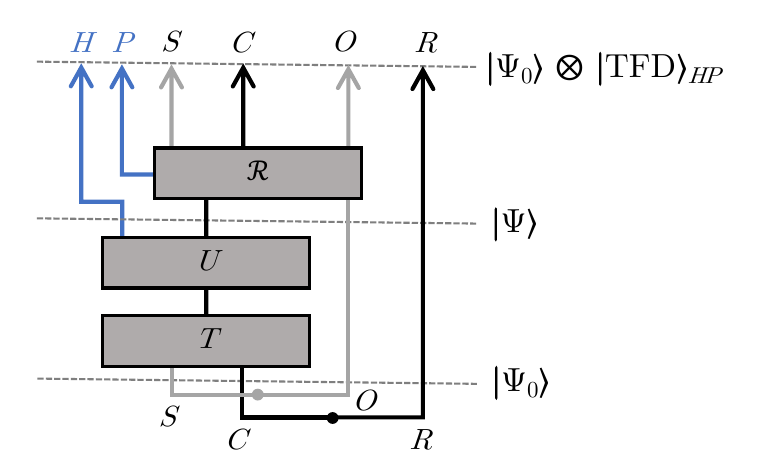}
  \caption{The system starts with $\ket{\Psi_0}$ shown in (\ref{Psi0}), which contains the code subspace $C$, early radiation $R$, soft hair $S$, and exterior observer $O$. The operation $T$ maps $S$ and $C$ into a black hole state, which then emits a Hawking radiation $H$ after the time evolution $U$. Finally, the observer recovers the Hawking partner $P$ using the recovery map $\mR$. The whole process is unitary.}\label{recover}
  \end{figure}

We start with the system $\ket{\Psi_0}$ shown in Fig. \ref{recover}, where we have the code subspace of the black hole $C$ correlated with the early radiation $R$, and the soft hair $S$ of the black hole correlated with the observer $O$. The corresponding Penrose diagram is shown in Fig. \ref{penrose}.
We can write the state $\ket{\Psi_0}$ as
\be
\ket{\Psi_0}=\frac{1}{\sqrt{d_C d_S}}\sum_{f,n}\ket{f}_S\ket{n}_C\ket{f}_O\ket{n}_R\,.\label{Psi0}
\ee
The operation $T$ embeds $S$ and $C$ onto a black hole state, which can be represented as
\be
T\ket{f}_S\ket{n}_C =T_f\ket{n}_B\,.
\ee
We can also define the inverse mapping $T_f^{\dagger}$ that reverse the above operation, which will be useful later.
Then, the black hole emits an extra Hawking radiation $H$ under the time evolution $U$. The system can be expressed as
\bea
\ket{\Psi} &=& U\circ T\ket{\Psi_0}\nn\\
&=&\frac{1}{\sqrt{d_C d_S}}\sum_{m,n,f}C_m\ket{m}_H\otimes T_f\ket{n}_B\ket{f}_O\ket{n}_R\,,
\eea
where $\ket{m}_H$ labels the Hawking radiation $H$, and $C_m$ is the Kraus operator. The recovery map $\mR$ is expected to recover the Hawking partner $P$ when the observer $O$ is included. We can call the process involving $O$ ``measurement". The recovery map $\mR$ maps the whole system into the original state $\ket{\Psi_0}$ and a thermo-field double (TFD) state composing $P$ and $H$, i.e. $\ket{\text{TFD}}_{HP}$. The process can be represented as
\bea
\mathcal{R} \ket{\Psi} &\simeq & \ket{\Psi_0} \otimes  \ket{\text{TFD}}_{HP}\nn\\
&=& \frac{1}{\sqrt{d_C d_S}}\sum_{f,n}\ket{f}_S\ket{n}_C\ket{f}_O\ket{n}_R \nn\\
&~&~~~~~~~~~~~~ \otimes \sum_m\sqrt{p_n}\ket{m}_P\ket{m}_H\,,
\eea
with the Boltzmann weights $p_n$, where we have used $\simeq$ because there is an error in the recovery map. It was shown that the Petz map can be used to accomplish the recovery task \cite{Pasterski2020}
\be
\mR =\frac{1}{\sqrt{d_C}}T^{\dagger}C^{\dagger}_m\sigma^{-1/2}\,~ \text{with}~
\sigma=\frac{1}{d_C}\sum_m C_m T_f T_f^{\dagger}C_m^{\dagger}\,.\nonumber
\ee

Now, by including the Hilbert space of the observer $O$, the decoupling of the early radiation $R$ and late time radiation $H$ and the reconstruction of Hawking partner $P$ are realized; thus, the AMPS firewall can be avoided through the above argument. The argument can be summarized as the Yoshida's decoupling theorem \cite{Yoshida2019} shown in section \ref{intro}. The mechanism of how soft hair interacts with Hawking radiation was recently explored \cite{Flanagan2021,Flanagan2021a}.
In paper \cite{Cheng2020}, the phase spaces of the Hawking radiation and the measurement were evaluated, and the competition between those two processes gives out the Page curve that is consistent with the unitarity.

\section{Fidelity of the reconstruction}
\label{fidelity}

Let us sketch a proof of the decoupling theorem, from which we can see more details about the fidelity of the reconstruction. To prove the decoupling theorem, we need to show that the density matrix $\rho_{HR}$ can be written as
\be
\rho_{HR}\approx \rho_{H}\otimes \rho_R\,,
\ee
for a scrambling evolution. We are going to show that $\rho_{HR}$ is very close to a maximally mixed state in $L^1$ distance, namely,
\be
||\rho_{HR}-\frac{1}{d_Hd_R}I_{H}\otimes I_{R}||_1^2\ll 1\,,
\ee
with identity matrices $I_H$ and $I_R$. It is easy to show that
\bea
||\rho_{HR}-\frac{1}{d_Hd_R}I_{H}\otimes I_{R}||_1^2&\leq&  ||\rho_{HR}-\frac{1}{d_Hd_R}I_{H}\otimes I_{R}||_2^2\nn\\
&=&d_Hd_R~\tr( \rho_{HR}^2)-1\,.\label{L1}
\eea
Then, the main task left for us is to evaluate the 2nd Renyi entropy, which can be done following the diagram technique \cite{Yoshida2017,Yoshida2019}.
Here, we will omit map $T$ shown in Fig. \ref{recover}, which maps the code subspace $C$ and soft hair $S$ to a black hole state, and only focus on the unitary evolution $U$. The density matrix $\rho_{HR}$ can be expressed diagrammatically as
\be
\rho_{HR}=\tr_{BO} \ket{\Psi}\bra{\Psi}=~
\begin{matrix}
\includegraphics[width=4cm]{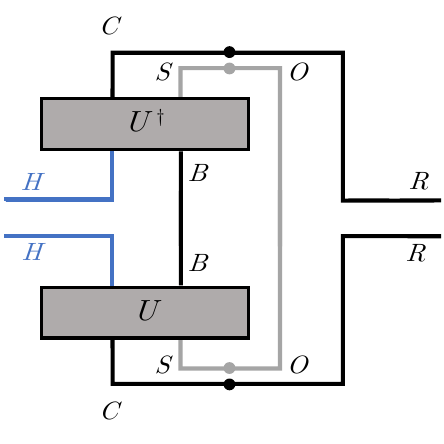}
\end{matrix}\label{rhoHR}
\ee
where we get the reduced density matrix by tracing out the Hilbert space of the observer $O$ and black hole $B$. Naturally, we can also illustrate $\tr (\rho_{HR}^2)$ as follows
\be
\tr (\rho_{HR}^2)=
\begin{matrix}
\includegraphics[width=4cm]{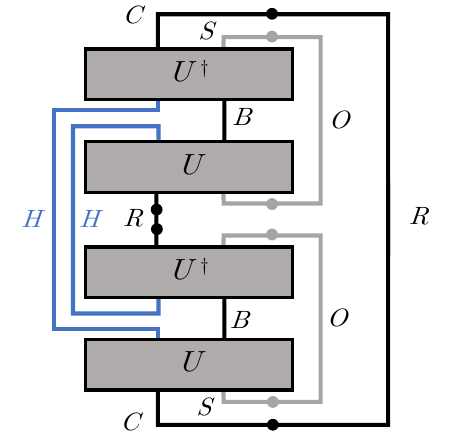}
\end{matrix}\label{rho2}
\ee
Note that inspired by the recent discussions of replica wormholes \cite{Penington2019a,Almheiri2019b}, there might be a different combination of the blocks and lines corresponding to a different saddle in Euclidean path integral, which requires further understanding. Here we would only focus on the above diagram shown in (\ref{rho2}). To evaluate $\tr (\rho_{HR}^2)$, we need to recall some properties of scrambling evolution. Quantum scrambling is related to the out-of-time-order correlators (OTOCs), and one of the characteristics of almost perfect scrambling can be expressed as the following decomposition \cite{Hosur2016,Yoshida2017}
\bea
&\left<  U^\dagger W U~Z~U^\dagger Y U~X \right> \nn\\
&\approx \left< WY\right>\left< Z\right>\left< X\right>+ \left< ZX\right>\left< W\right>\left< Y\right>-\left< W\right>\left< X\right>\left< Y\right>\left< X\right>\,,\nn\\
\eea
where $U$ is a scrambling evolution. Now, with four $U$s in (\ref{rho2}), $\tr (\rho_{HR}^2)$ can also be expressed as an OTOC of some operators, which can be evaluated applying the above decomposition. We can get rid of $U$s by applying the decomposition and directly calculate the correlators, which are just some circles contracted with themselves, diagrammatically. So for the perfect scrambling evolution $U$, we have \cite{Yoshida2017,Yoshida2019}
\bea
\tr(\rho_{HR}^2)&\approx& \frac{d_H}{d_R}(\frac{1}{d_H^2}+\frac{1}{d_S^2}-\frac{1}{d_H^2d_S^2})\nn\\ &\leq& \frac{1}{d_Hd_R}+\frac{d_H}{d_Rd_S^2}\,.\label{trresult}
\eea
substituting the above result (\ref{trresult}) back into the $L^1$ distance (\ref{L1}), we obtain
\be
||\rho_{HR}-\frac{1}{d_Hd_R}I_{H}\otimes I_{R}||_1^2\sim \mO\left(\frac{d_H^2}{d_S^2}\right)\,,
\ee 
and the fidelity $F$ always satisfies the following inequality
\be
F \leq 1-\mO\left(\frac{d_H^2}{d_S^2}\right)\,.
\ee
So in order to keep a high fidelity, $d_S$ must be much larger than $d_H$ such that
\be
\mO\left(\frac{d_H^2}{d_S^2}\right)\ll 1\,.
\ee

\section{Hair-loss paradox}
\label{hair-loss}

\begin{figure}
  \centering
\includegraphics[width=8cm]{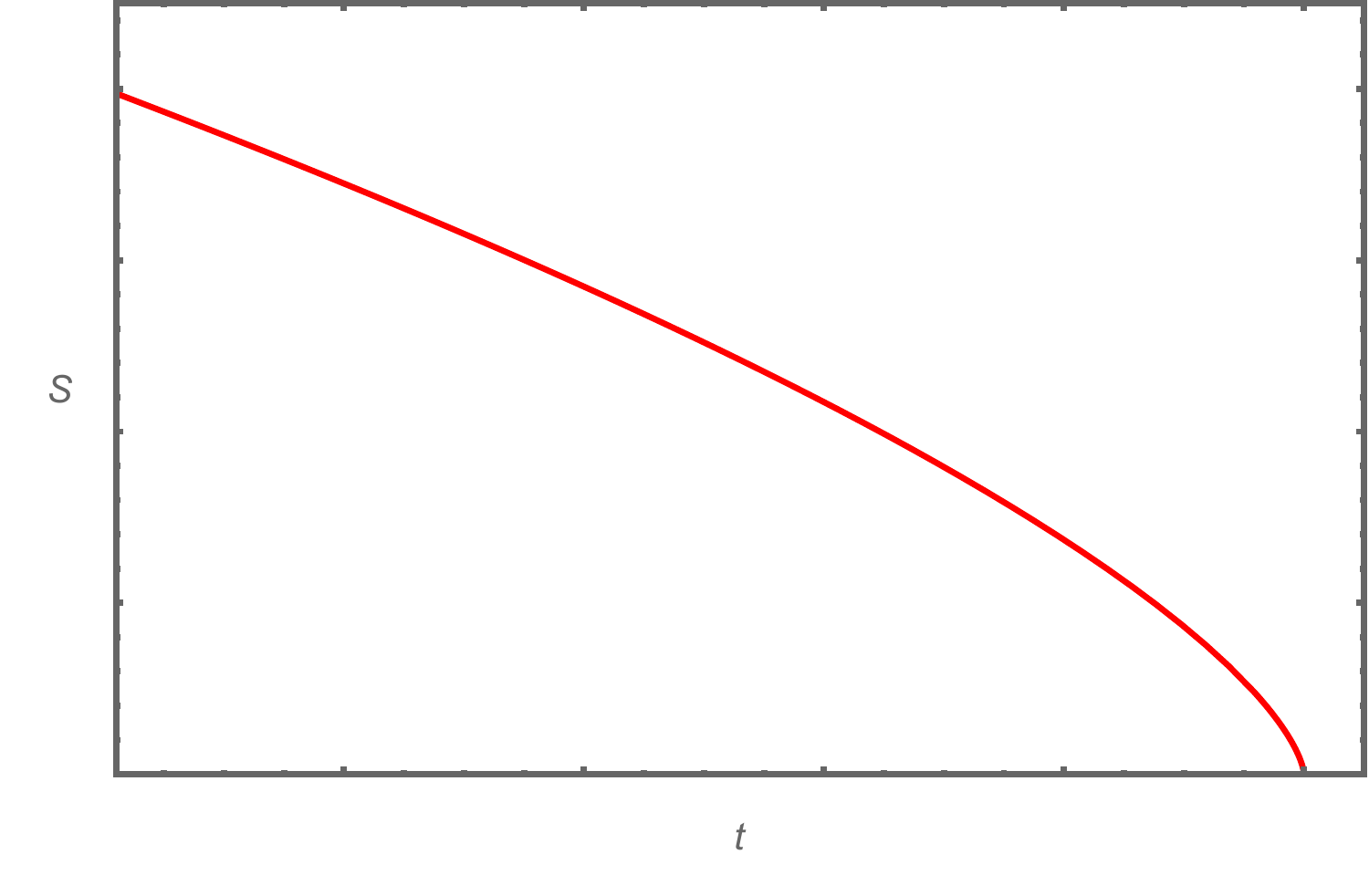}
  \caption{The coarse-grained entropy of an evaporating black hole decreases to zero.}\label{Aparadox}
\end{figure}

From the previous section, we have gotten an important condition to ensure the decoupling theorem to be true. Supposing the correlations between the extra system $O$ and the black hole are through soft hair \cite{Pasterski2020}, the dimension of the soft hair Hilbert space $\mH_S$, which equals the dimension of the exterior observer, must be very large compared to the late time Hawking radiation $\mH_H$, and the error is proportional to
\be
\text{Error}\sim \mO\left(\frac{d_H^2}{d_S^2}\right)\ll 1\,.\label{169}
\ee
So to properly complete the task, $d_S$ must be much larger than $d_H$. This condition will bring us another paradox in the process of black hole evaporation.

Soft hair can at most be responsible for part of the black hole micro-states, and the coarse-grained entropy of soft hair should always be smaller than the coarse-grained entropy of the black hole. So the soft hair Hilbert space $\mH_S$ can never be larger than the whole Hilbert space of the black hole; thus, we have
\be
d_S\leq d_{BH}\,.\label{170}
\ee
The combination of equations (\ref{169}) and (\ref{170}) implies that in order to decouple the late and early radiation, and reconstruct the interior operators with a high fidelity, the following inequality.
\be
d_{BH}\gg d_H\,\label{condition}
\ee
must be true. However, as the black hole evaporates, $d_{BH}$ is decreasing. The decoupling theorem fails to hold once the above condition can not be fulfilled and the firewall comes back again. In such a sense, we have not completely wiped out the firewall at the Page time but postponed it until the end stage of the evaporation.
So, we can say that we keep losing soft hair degrees of freedom as the black hole evaporates because the entropy of the soft hair must be smaller than the coarse-grained entropy of the black hole. 
When we do not have enough soft hair, the decoupling theorem fails. So we call it

\noindent \textbf{Hair-loss paradox:} \textit{At the end stage of the black hole evaporation, the black hole system does not have large enough coarse-grained entropy to support large $\mH_S$ to decouple the early and late radiation and construct the Hawking partner.}

The above hair-loss paradox is a paradox in a more general setting than a paradox that only appears in Yoshida's decoupling theorem. Most of the attempts trying to reconstruct the interior operators using quantum information protocol rely on a large phase space of the radiation. If the black hole system is only left with a few qubits, even still exponentially hard, the HHA decoding task \cite{Harlow2013, Aaronson2016} is not that hard. Thus, there is no decoupling between the early and late radiation, and we always have to face the AMPS firewall if the black hole system only has several qubits.
If the firewall shows up once on the horizon, there would be a firewall all over the horizon because of the general covariance of the general relativity.
This paradox can be seen more explicitly by looking at the entropy of an evaporating black hole \cite{Page1993, Page2013}. The Stephan-Boltzmann law of blackbody radiation can be written as
\be
\frac{dM}{dAdt}\propto T^4,
\ee
with the area of the black hole $A$. Then, the power of the evaporation can be written as
\be
\frac{dM}{dt}\propto \frac{1}{M^2}.
\ee
Working out the above integral, we obtain
\be
M=M_0\left(1-\frac{t}{t_0}\right)^{1/3}\,,
\ee
where $M_0$ is the original mass of the black hole, and $t_0$ is the total time for the evaporation. Similarly, for the system's entropy, we have
\be
S=S_0\left(1-\frac{t}{t_0}\right)^{2/3}\,,
\ee
with the original entropy of the black hole $S_0$. The coarse-grained entropy of the black hole system is depicted in Fig. \ref{Aparadox}. As we can see from the figure, at the end of the evaporation, the coarse-grained entropy of the system goes to zero, thus can not support a large number of soft hair degrees of freedom for the purpose of decoupling between the late and early time radiation and constructing the Hawking partner.

It is important to notice that the hair-loss paradox was also hinted in the classical gravity when considering a large density firewall \cite{Abramowicz2013}. In Appendix \ref{classical}, we checked that the firewall in our work is compatible with Einstein’s theory of gravity, while the firewall appears near the Page time is not.

\section{Summary}

We have presented a paradox related to the decoupling theorem of evaporating black holes. The decoupling between the early and late Hawking radiation can be realized by including soft hair in a black hole background. The interior operator can also be reconstructed by the Petz map, without using the early radiation $R$. What's more, for scrambling evolution $U$, we can evaluate the fidelity of the decoupling between early and late Hawking radiation. However, in order to keep a high fidelity, the dimension of the soft hair Hilbert space $d_S$ must be much larger than the one for late Hawking radiation $d_H$. We argue that the condition for a high fidelity might not hold at the end stage of the evaporation, which means the firewall would show up again even if the decoupling theorem holds near the Page time. The paradox is quite common in most schemes trying to decouple the late and early radiation to avoid firewall, for example the HHA decoding task.

A possible way of avoiding this problem is that there might be a large modification of the black hole entropy at the end stage of the evaporation, such that the condition (\ref{condition}) is always held. This has a strong flavor of the black hole remnants \cite{Aharonov1987, Adler2001}. 
However, a more precise solution to the hair-loss paradox needs further studies. A better understanding of the end-stage of evaporation relies on a quantum theory of gravity, which we don't know much about. Maybe now is the time to consider more about the end of the evaporation, which may give us tons of new insights.

\begin{acknowledgments}
We would like to thank Yang An, Jan de Boer, Diego Hofman, Ran Li, Pujian Mao, Shao-Wen Wei for inspirational discussions. We thank the referees for critical remarks that helped us a lot.
PC is financially supported by China Scholarship Council (CSC).
\end{acknowledgments}

\appendix

\section{Consistency with classical gravity}
\label{classical}

In this appendix, we are going to show that the firewall discussed in this \LorP~ is compatible with Einstein's theory of gravity. 

Soon after the AMPS firewall was put forward, Abramowicz, Kluzniak, and Lasota \cite{Abramowicz2013} checked the possible conflict between a Planck density firewall and Einstein's classical gravity. It was shown that the Planck density firewall near the horizon of a large-mass black hole is not compatible with classical gravity, and the classical gravity implies an upper limit of the total mass of the black hole.
Let us denote the upper mass limit as $M_{up}$. The upper limit $M_{up}$ is of order Planck mass $M_{pl}$, which is very tiny compared to astrophysical black holes. 

The upper limit can be loosened by decreasing the density of the firewall. Let us restrict our attention to a static observer. If the firewall density is reduced by a factor of $\beta$, the upper mass limit of the black hole can be rewritten as \cite{Abramowicz2013}
\be
M_{up}=\frac{M_{pl}}{8\pi\beta}\,.
\ee 
Now, let's suppose the energy scale of the firewall is GUT (Grand Unified Theory) energy $10^{16}$ GeV\footnote{Note that the energy scale is high enough to destroy any in-falling particles. To get an impression, the highest energy scale human can reach is on the LHC (Large Hadron Collider) at CERN, which is about $10^4$ GeV.  $10^{16}$ GeV is about one trillion times higher than LHC. }, then the upper limit of the black hole total mass is 
\be
M_{up}\sim 10^3\times M_{pl}\,.
\ee
So, the conclusion is very straightforward, which is that a very high-energy firewall cannot arise for astrophysical black holes, but can appear for black holes of thousands of Planck mass. It is obvious that the firewall discussed in the current \LorP~ satisfies the above condition. In such a sense, the firewall appears at the end of the evaporation, which is indicated in the \LorP, is the possible firewall allowed by Einstein's equations. 
 
In such a sense, the firewall appears at the end of the evaporation, which is indicated in the current \LorP, is the possible firewall allowed by Einstein's equations. 
However, it is important to address the following point. Saying that the firewall at the end of the evaporation is compatible with the classical gravity doesn't mean that we should accept the firewall. The appearance of the firewall is always a violation of the smoothness of the near horizon geometry, and the violation of the monogamy of entanglement is still a puzzle to solve as mentioned in section \ref{intro}. The point we are trying to make is that even though the firewall might be avoided by adding some extra ingredients, like soft hair, the firewall can still reappear at the end of the evaporation, and a smarter scheme might be needed to better understand the physics at the end of the evaporation.


\begin{thebibliography}{10}

\bibitem{Hawking1976}
S.W.~Hawking, \emph{Breakdown of predictability in gravitational collapse},
  \href{https://doi.org/10.1103/physrevd.14.2460}{\emph{Phys Rev D} {\bfseries
  14} (1976) 2460}.

\bibitem{Almheiri2012}
A.~Almheiri, D.~Marolf, J.~Polchinski and J.~Sully, \emph{Black holes:
  Complementarity or firewalls?},
  \href{https://doi.org/10.1007/JHEP02(2013)062}{\emph{J High Energy Phys}
  {\bfseries 2013} (2013) } [\href{https://arxiv.org/abs/1207.3123}{{\ttfamily
  1207.3123}}].

\bibitem{Page1993}
D.N.~Page, \emph{Information in black hole radiation},
  \href{https://doi.org/10.1103/PhysRevLett.71.3743}{\emph{Phys Rev Lett}
  {\bfseries 71} (1993) 3743}
  [\href{https://arxiv.org/abs/hep-th/9306083v2}{{\ttfamily
  hep-th/9306083v2}}].

\bibitem{Page2013}
D.N.~Page, \emph{Time dependence of Hawking radiation entropy},
  \href{https://doi.org/10.1088/1475-7516/2013/09/028}{\emph{J Cosmol Astropart
  Phys} {\bfseries 2013} (2013) 028}
  [\href{https://arxiv.org/abs/1301.4995v3}{{\ttfamily 1301.4995v3}}].

\bibitem{Terhal2004}
B.M.~Terhal, \emph{Is entanglement monogamous?},
  \href{https://doi.org/10.1147/rd.481.0071}{\emph{{IBM} Journal of Research
  and Development} {\bfseries 48} (2004) 71}.

\bibitem{Harlow2013}
D.~Harlow and P.~Hayden, \emph{Quantum computation vs. firewalls},
  \href{https://doi.org/10.1007/JHEP06(2013)085}{\emph{J. High Energy Phys.}
  {\bfseries 2013} (2013) } [\href{https://arxiv.org/abs/1301.4504}{{\ttfamily
  1301.4504}}].

\bibitem{Aaronson2016}
S.~Aaronson, \emph{The complexity of quantum states and transformations: From
  quantum money to black holes},
  \href{https://arxiv.org/abs/1607.05256}{{\ttfamily 1607.05256}}.

\bibitem{Susskind_1993}
L.~Susskind, L.~Thorlacius and J.~Uglum, \emph{The stretched horizon and black
  hole complementarity},
  \href{https://doi.org/10.1103/physrevd.48.3743}{\emph{Physical Review D}
  {\bfseries 48} (1993) 3743}
  [\href{https://arxiv.org/abs/hep-th/9306069}{{\ttfamily hep-th/9306069}}].

\bibitem{Kiem1995}
Y.~Kiem, E.~Verlinde and H.~Verlinde, \emph{Black hole horizons and
  complementarity},
  \href{https://doi.org/10.1103/PhysRevD.52.7053}{\emph{Phys.Rev. D52 (1995)
  7053-7065} (1995) } [\href{https://arxiv.org/abs/hep-th/9502074}{{\ttfamily
  hep-th/9502074}}].

\bibitem{Lowe1995}
D.A.~Lowe, J.~Polchinski, L.~Susskind, L.~Thorlacius and J.~Uglum, \emph{Black
  hole complementarity vs. locality},
  \href{https://doi.org/10.1103/PhysRevD.52.6997}{\emph{Phys.Rev.D} {\bfseries 52}
  (1995) 6997} [\href{https://arxiv.org/abs/hep-th/9506138}{{\ttfamily
  hep-th/9506138}}].

\bibitem{Yoshida2019}
B.~Yoshida, \emph{Observer-dependent black hole interior from operator collision}, \href{https://link.aps.org/doi/10.1103/PhysRevD.103.046004}{\emph{Phys. Rev. D} {\bfseries 103} (2021) 046004}
  [\href{https://arxiv.org/abs/1910.11346}{{\ttfamily 1910.11346}}].

\bibitem{Yoshida2017}
B.~Yoshida and A.~Kitaev, \emph{Efficient decoding for the Hayden-Preskill
  protocol},  \href{https://arxiv.org/abs/1710.03363}{{\ttfamily 1710.03363}}.

\bibitem{Hosur2016}
P.~Hosur, X.-L.~Qi, D.A.~Roberts and B.~Yoshida, \emph{Chaos in quantum
  channels}, \href{https://doi.org/10.1007/jhep02(2016)004}{\emph{Journal of
  High Energy Physics} {\bfseries 2016} (2016) }.

\bibitem{Yoshida2019a}
B.~Yoshida, \emph{Soft mode and interior operator in the Hayden-Preskill
  thought experiment},
  \href{https://link.aps.org/doi/10.1103/PhysRevD.100.086001}{\emph{Phys. Rev. D}
  {\bfseries 100} (2019) 086001}.

\bibitem{Pasterski2020}
S.~Pasterski and H.~Verlinde, \emph{HPS meets AMPS: How soft hair dissolves the
  firewall},  \href{https://arxiv.org/abs/2012.03850}{{\ttfamily 2012.03850}}.

\bibitem{Petz1988}
D.~Petz, \emph{Sufficiency of channels over von Neumann algebras}, \href{https://doi.org/10.1093/qmath/39.1.97}{\emph{Q. J. Math.}
  {\bfseries 39} (1988) 97}.

\bibitem{DenesPetz2004}
M.O.~Denes~Petz, \emph{Quantum entropy and its use}, Springer Berlin Heidelberg
  (Mar., 2004).

\bibitem{Cheng2020}
P.~Cheng and Y.~An, \emph{Soft black hole information paradox: Page curve from
  Maxwell soft hair of a black hole},
  \href{https://doi.org/10.1103/PhysRevD.103.126020}{\emph{Phys. Rev. D}
  {\bfseries 103} (2021) 126020}.
  
\bibitem{Li2021a}
R.~Li and J.~Wang, \emph{Hayden-Preskill protocol and decoding Hawking
  radiation at finite temperature},
  \href{https://arxiv.org/abs/2108.09144}{{\ttfamily 2108.09144}}.
  
  
\bibitem{Himwich2020}
E.~Himwich, S.A.~Narayanan, M.~Pate, N.~Paul and A.~Strominger, \emph{The soft
  $\mathcal{S}$-matrix in gravity}, \href{https://doi.org/10.1007/JHEP09(2020)129}{\emph{J
  High Energy Phys} {\bfseries 2020} (2020) }
  [\href{https://arxiv.org/abs/2005.13433}{{\ttfamily 2005.13433}}].

\bibitem{Strominger2017}
A.~Strominger, \emph{Lectures on the infrared structure of gravity and gauge
  theory},  \href{https://arxiv.org/abs/1703.05448v2}{{\ttfamily
  1703.05448v2}}.

\bibitem{Donnay2016}
L.~Donnay, G. Giribet, H. A.~Gonz\'alez, and M.~Pino, \emph{Supertranslations and Superrotations at the Black Hole Horizon}, 
\href{https://doi.org/10.1103/PhysRevLett.116.091101}{ \emph{Phys. Rev. Lett. }{\bfseries 116}, 091101 (2016)}
  [\href{https://arxiv.org/abs/1511.08687}{{\ttfamily 1511.08687}}].


\bibitem{Hawking2016}
S.W.~Hawking and M.J.~Perry and A.~ Strominger, \emph{Soft hair on black holes}, 
\href{https://link.aps.org/doi/10.1103/PhysRevLett.116.231301}{ \emph{Phys. Rev. Lett. }{\bfseries 116}, 231301 (2016)}
  [\href{https://arxiv.org/abs/1601.00921}{{\ttfamily 1601.00921}}].


\bibitem{Cheng2022}
P.~Cheng and P.~Mao, \emph{Soft theorems in curved spacetime}, to appear.

\bibitem{Flanagan2021}
{\'{E}}.{\'{E}}.~Flanagan, \emph{Order-unity correction to Hawking radiation},
  \href{https://doi.org/10.1103/physrevlett.127.041301}{\emph{Physical Review
  Letters} {\bfseries 127} (2021) 041301}.

\bibitem{Flanagan2021a}
{\'{E}}.{\'{E}}.~Flanagan, \emph{Infrared effects in the late stages of black
  hole evaporation},
  \href{https://doi.org/10.1007/jhep07(2021)137}{\emph{Journal of High Energy
  Physics} {\bfseries 2021} (2021) }.

\bibitem{Penington2019a}
G.~Penington, S.H.~Shenker, D.~Stanford and Z.~Yang, \emph{Replica wormholes
  and the black hole interior},
  \href{https://arxiv.org/abs/1911.11977v2}{{\ttfamily 1911.11977v2}}.

\bibitem{Almheiri2019b}
A.~Almheiri, T.~Hartman, J.~Maldacena, E.~Shaghoulian and A.~Tajdini,
  \emph{Replica wormholes and the entropy of Hawking radiation},
  \href{https://doi.org/10.1007/JHEP05(2020)013}{\emph{J High Energy Phys}
  {\bfseries 2020} (2020) }
  [\href{https://arxiv.org/abs/1911.12333v2}{{\ttfamily 1911.12333v2}}].

\bibitem{Abramowicz2013}
M.~A.~Abramowicz, W.~Klu\'zniak and J.~P.~Lasota,
\emph{Mass of a Black Hole Firewall},
\href{http://doi.org/10.1103/PhysRevLett.112.091301}{Phys. Rev. Lett. \textbf{112} (2014) 091301}
[\href{https://arxiv.org/abs/1311.0239}{{\ttfamily 1311.0239}}].


\bibitem{Aharonov1987}
Y.~Aharonov, A.~Casher and S.~Nussinov, \emph{The unitarity puzzle and Planck
  mass stable particles},
  \href{https://doi.org/10.1016/0370-2693(87)91320-7}{\emph{Physics Letters B}
  {\bfseries 191} (1987) 51}.

\bibitem{Adler2001}
R.J.~Adler, P.~Chen and D.I.~Santiago, \emph{The generalized uncertainty
  principle and black hole remnants},
  \href{https://doi.org/10.1023/a:1015281430411}{\emph{General Relativity and
  Gravitation} {\bfseries 33} (2001) 2101}.

\end{thebibliography}
\bibliographystyle{./global/JHEP}

\providecommand{\href}[2]{#2}
\begingroup\raggedright

\endgroup

\end{document}